\documentclass[16pt]{article}
\usepackage{soul}

\usepackage[square,numbers,sort&compress]{natbib}
\usepackage{graphicx} 
\usepackage{amsmath,graphicx,mathtools,setspace,subfigure}
\usepackage[colorlinks=False]{hyperref}
\hypersetup{
  colorlinks=false,
  linkbordercolor={0 0 0.45},
  citebordercolor={0 0 0.45},
  urlbordercolor={0 0 0.45}
}
\usepackage{bm}
\usepackage{authblk}
\usepackage{amsfonts}
\usepackage{color}
\usepackage{amsthm}
\usepackage{mathrsfs}
\usepackage{bm}
\usepackage{comment}
\usepackage{braket}
\usepackage{ulem}
\usepackage[export]{adjustbox}

\def\m{M$^3$}
\def\ec{EC$^3$}
\def\cw{${\cal L}w_{1+\infty}$}
\def\dw{$ {\cal L}_{\Lambda}w_{1+\infty}$}

\def\half{ \frac{1}{2} }
\def\ee{\end{equation}}
\def\be{\begin{equation}}
\def\bea{\begin{eqnarray}}
\def\eea{\end{eqnarray}}

\begin{document}
\begin{titlepage}
\unitlength = 1mm~\\
\vskip 3cm
\begin{center}

{\LARGE{EVERY  CFT$_3$ HAS AN  \dw\  SYMMETRY}}

\vspace{0.8cm}
Andrew Strominger and Hongji Wei \\
\vspace{1cm}

{\it  Center for the Fundamental Laws of Nature, Harvard University, Cambridge, MA 02138, USA} \\

\vspace{0.8cm}

\begin{abstract}
Recently a one-parameter family of deformed \cw\ soft symmetry algebras, denoted \dw ,   acting on tree-level gravitational theories in 
AdS$_4$ has been discovered. Here we show that all  CFT$_3$s, including those dual to quantum gravity on AdS$_4$,  admit an \dw\ action generated by the 
ANEC operator, its conformal descendants and their commutators. This extends the previous tree-level results on these soft  symmetries  to the strongly-coupled  quantum regime.  
 \end{abstract}

\end{center}

\end{titlepage}

\tableofcontents

\section{Introduction}
Nonabelian gauge theory with gauge group $\mathcal{G}$ in Minkowski space  has an infinite-dimensional asymptotic symmetry algebra known as the $S$-algebra. The algebra  is  generated by soft gluons at tree level \cite{Guevara:2021abz,Strominger:2021mtt}.  Recently \cite{Sheta:2025oep} it has been shown that the  soft  gluons  can be conformally mapped to AdS$_4$, where they generate the same $S$-algebra. In the CFT$_3$ dual to AdS$_4$, the soft gluons are realized as light transforms of the conserved global $\mathcal{G}$ current,  its   conformal descendants and their commutators. The fact that this family of  light ray operators obey the $S$-algebra can be shown directly within any CFT$_3$ with a conserved $\mathcal{G}$ current without reference to the bulk dual.  

One seeks a  generalization of  this relation between flat space and AdS$_4$ soft symmetries  to gravity, where the flat space soft  symmetry group is  \cw ~\cite{Strominger:2021mtt,Ball:2021tmb,mason2023,adamo2022,kmec2024,himwich2023,Miller:2025wpq,Bu:2024wnf,Ruzziconi:2025fuy,Cresto:2024fhd,Donnay:2024qwq}. At first, this may sound problematic because gravity is not conformally invariant. However, in beautiful recent work \cite{Schwarz2023SymmetriesCelestial,taylor2023, bittleston2024,Jorge2023} a deformed algebra was discovered which  acts\footnote{Modulo boundary conditions, see section 5.} on tree-level Einstein gravity in AdS$_4$ (or dS$_4$) with cosmological constant $\Lambda$. The deformed algebra, denoted  \dw,  is:  
\begin{align}
    \left[w^p_{\bar{m},m},w^q_{\bar{n},n}\right] = (\bar{m}(q-1)-\bar{n}(p-1))w^{p+q-2}_{\bar{m}+\bar{n},m+n} -\Lambda   (m(q-2)-n(p-2))w^{p+q-1}_{\bar{m}+\bar{n},m+n}.
\end{align}
The existence of a deformation of \cw\  obeying the Jacobi identity and with a global $SO(3,2)$ AdS$_4$ isometry group is remarkable. It suggests there may after all, despite the absence of the usual type of soft theorems,  be a  map from the flat space \cw\ generators to \dw\ generators in AdS$_4$, in which the deformation of  the algebra arises because gravity is not conformally invariant.

In this paper we establish the existence of a \dw\ symmetry directly in CFT$_3$. This   provides the  dual realization of the bulk  \dw\ symmetry in AdS$_4$ and generalizes \cite{Sheta:2025oep} to gravity.   We work on the $S^2\times {\mathcal R}$ Einstein cylinder EC$^3$. Every CFT$_3$ has an ANEC light ray operator given by  the integral of the stress tensor over a null line beginning at an initial point $x_i$ and ending at the antipodal point in EC$^3$\cite{HartmanMathys2024LightRaySumRules,HartmanMathys2023ANECRG,FaulknerLeighParrikarWang2016ModularANEC,BalakrishnanFaulknerKhandkerWang2019QNEC,HartmanKunduTajdini2017ANECFromCausality,CasiniTesteTorroba2017NullPlaneMarkov,HofmanLiMeltzerPoland2016ColliderBounds,Kabat_2021,Rosso2020GlobalAspectsANEC,moult2025memorycorrelatorswardidentities,Gonzalez:2025ene}.  There is a one-parameter family of such operators labeled by the angle at which the null line emanates from $x_i$. We consider all such light ray operators, their conformal descendants and their commutators.  We show that the  
commutator algebra of these operators is \dw, with a wedge restriction on the generators given in Section 4. Our conclusion follows from known results about light ray operators and their commutators in the CFT literature \cite{Cordova_2018,Himwich:2025ekg,Hofman_2008,Kravchuk_2018,Karateev_2018, Kologlu:2019mfz,Korchemsky:2021htm,BelinHofmanMathysWalters2021StressTensorLightRay,BeskenDeBoerMathys2021IntegratedCommutators,ChangKologluKravchukSimmonsDuffinZhiboedov2022TransverseSpin,Huang2019NearLightcone,Huang_2020,Huang2021NearLine,HuPasterski2023CelestialConformalColliders,HuPasterski2023DetectorOperators,GonzoPokraka2021Detectors,Sharma2022AmbidextrousLightTransforms,BanerjeeBasuBhatkar2023LightTransformedGluon,HuEtAl2022FourPointLightRay,Korchemsky_2022,DeHuYelleshpurVolovich2022FourLightRay,BalakrishnanChandrasekaranFaulknerLevineShahbaziMoghaddam2022ReplicaDefects,FreivogelStoffels2025PositivityLightRay}. Our analysis is largely  from the boundary point of view. A bulk analysis like the one given for gauge theory in \cite{Sheta:2025oep} would be of interest. 

This extends the perturbative tree-level results of \cite{taylor2023,bittleston2024,Jorge2023} and establishes that \dw\ can act on strongly interacting quantum systems. 

An  interpretation of the \dw\ algebra for gravity in   the positive $\Lambda$ de Sitter case remains an open problem. 

\section{Cordova-Shao Light Ray Operators}

In reference~\cite{Cordova_2018}, Cordova and Shao studied the following CFT$_3$ light ray operators\footnote{Throughout  this paper we specialize their $D$-dimensional results to $D=3$.}
\begin{align}\label{eqn: lightray E flat}
    \mathcal{E}(y) &=  \int_{-\infty}^{\infty} dy^+  T_{++}(y^+,0,y), \\
   \mathcal{K}(y)&= \int_{-\infty}^{\infty} dy^+  y^+T_{++}(y^+,0,y),\label{eqn: lightray K flat}\\
    \mathcal{N}(y)& = \int_{-\infty}^{\infty}dy^+ T_{+y}(y^+,0,y) .\label{eqn: lightray N flat}
\end{align}
where $T$ is the stress tensor,  the \m\ (3D Minkowski) metric is $ds_{\text{M}^3}^2=-dy^+dy^-+dy^2$, and all operators are at $y^-=0$. 
\eqref{eqn: lightray E flat} is the ANEC operator, while \eqref{eqn: lightray K flat},\eqref{eqn: lightray N flat} are generalizations.

\begin{figure}
    \centering
    \includegraphics[width=55mm]{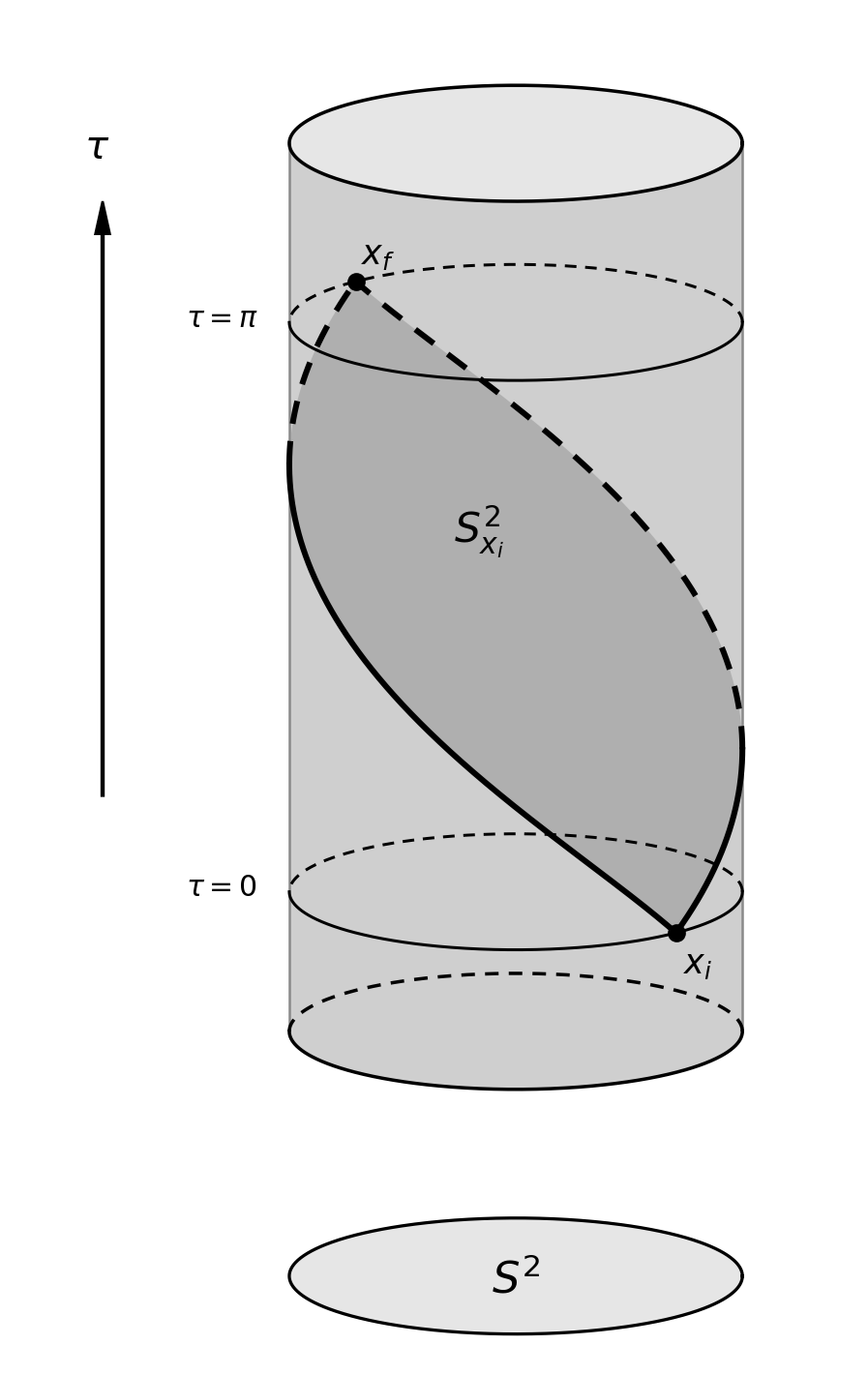}
    \caption{The dark grey region depicts the  set of all light rays in the  $S^2\times R$ Einstein cylinder EC$^3$  (forming the boundary of AdS$_4$) beginning at an initial  point $x_i$ and reconverging  at the antipodal  point $x_f$. These comprise a null $S^2$ and are a Cauchy surface for EC$^3$.  The sphere $S^2$ is shown schematically as a disk.}
    \label{fig: EC}
\end{figure}

We wish to study these and other light ray operators  on the Einstein cylinder (\ec) with metric 
\bea\label{eqn: EC null metric}
    ds^2_{\text{EC}^3} &=& -d\tau^2+d\theta^2+\sin^2\theta d\phi^2 \cr &=&- 4d\tau^+ d\tau^- + \sin^2(\tau^+-\tau^-) d\phi^2,~~~~~\tau^\pm={\tau \pm \theta \over 2}.
\eea
The conformal map from \m\ to \ec\ is given in the appendix \ref{appendix: Coords Light Ray}. The transformation properties of light ray operators under this map has been studied in \cite{Hofman_2008}.
One finds 
\begin{align}\label{eqn:lightray E EC}
    \mathcal{E}(\phi) &= \frac{1}{2} \int_0^\pi d\tau^+ \sin^3 \tau^+ T_{++}(\tau^+,0,\phi), \\
   \mathcal{K}(\phi)&= -\int_0^\pi d\tau^+ \sin^2 \tau^+ \cos \tau^+ T_{++}(\tau^+,0,\phi),\label{eqn: lightray K EC}\\
    \mathcal{N}(\phi)& =  \hspace{7pt}\int_0^\pi d\tau^+ \sin \tau^+ T_{+\phi}(\tau^+,0,\phi) ,\label{eqn: lightray N EC}
\end{align}
at $\tau^-=0$. Note that on \ec\ all the line integrals begin at the same point $x_i=(0,0,\phi)$ and end at the same antipodal point $x_f=(\pi,\pi,\phi)$. We denote the space of lines as $\phi$ varies from $0$ to $2\pi$, depicted in Fig.~\ref{fig: EC},  by  $S^2_{x_i}$. It is a segment of the future light cone of $x_i$ and comprises a Cauchy surface in \ec. 

Finally it is useful to transform to a mode basis for the operators 
\begin{align}\label{mbss}
    \mathcal{E}_k &= \int_0^{2\pi}d\phi e^{ik\phi} \mathcal{E}(\phi)  \\
   \mathcal{K}_k&= \int_0^{2\pi}d\phi e^{ik\phi}\mathcal{K}(\phi)\\
    \mathcal{N}_k& =\int_0^{2\pi}d\phi e^{ik\phi} \mathcal{N}(\phi) .\end{align}
        
The commutators of the lightray operators \eqref{eqn: lightray E flat}-\eqref{eqn: lightray N flat} in flat coordinates were computed for any CFT$_3$ in \cite{Cordova_2018}. 
Conformally mapping to the \ec\ frame and Fourier transforming around the $\phi$ circle, their results can be rewritten
\begin{align}
    \left[\mathcal{E}_k,\mathcal{E}_l\right]&=0, \hspace{40pt} 
\left[\mathcal{K}_k,\mathcal{K}_l\right]=0,\hspace{40pt} \left[\mathcal{K}_k,\mathcal{E}_l\right] = -i \mathcal{E}_{k+l}, \label{eqn: comm E E K K K E}\\
\left[\mathcal{N}_k,\mathcal{E}_l\right]&=-l\mathcal{E}_{k+l},\hspace{10pt}
\left[\mathcal{N}_k,\mathcal{K}_l\right]=-l\mathcal{K}_{k+l},\hspace{10pt}
\left[\mathcal{N}_k,\mathcal{N}_l\right]  = (k-l) \mathcal{N}_{k+l},\label{eqn: comm N E N K N N}
\end{align}
\section{$ SO(3,2)$ Conformal Action}
The Cordova-Shao light ray operators \eqref{eqn: lightray E flat}-\eqref{eqn: lightray N flat} do not close under the action of the 3D conformal group $SO(3,2)$. In this section we construct the full representation of which they form a part, with some details deferred to Appendix \ref{app: every object}. 

$SO(3,2)$ is generated by 10 conformal Killing vectors (CKVs).
A convenient basis of CKVs is: 
\begin{align}
    L_{1} &=  \frac{i}{2}e^{ i\phi}\left[\cos^2 \tau^+\partial_+ - \cos^2\tau^-\partial_- +2i\cos \tau^+\cos \tau^-\csc(\tau^+-\tau^-) \partial_\phi\right]\\
     L_0 &=-\frac{i}{2}\left[\sin \tau^+ \!\cos \tau^+\partial_+ + 
    \sin \tau^- \!\cos \tau^- \partial_- + i\partial_\phi\right]\\
    L_{-1} &=  \frac{i}{2}e^{ -i\phi}\left[\sin^2 \tau^+\partial_+ - \sin^2\tau^-\partial_- -2i\sin \tau^+\sin \tau^-\csc(\tau^+-\tau^-) \partial_\phi\right]\\
 \bar{L}_{1} &=  \frac{i}{2}e^{-i\phi}\left[\cos^2 \tau^+\partial_+ - \cos^2\tau^-\partial_- -2i\cos \tau^+\cos \tau^-\csc(\tau^+-\tau^-) \partial_\phi\right]\\
    \bar{L}_0 &=-\frac{i}{2}\left[\sin \tau^+ \!\cos \tau^+\partial_+ + 
    \sin \tau^- \!\cos \tau^- \partial_- - i\partial_\phi\right]\\
    \bar{L}_{-1} &=  \frac{i}{2}e^{i\phi}\left[\sin^2 \tau^+\partial_+ - \sin^2\tau^-\partial_- +2i\sin \tau^+\sin \tau^-\csc(\tau^+-\tau^-) \partial_\phi\right]\\
    H_{\frac{1}{2},\frac{1}{2}} &= -\frac{1}{\sqrt{2}}\left[\cos^2 \tau^+\partial_+ + \cos^2 \tau^- \partial_-\right]\\
    H_{-\frac{1}{2},-\frac{1}{2}} &= -\frac{1}{\sqrt{2}}\left[\sin^2 \tau^+\partial_+ + \sin^2 \tau^- \partial_-\right]\\
    H_{\mp\frac{1}{2},\pm\frac{1}{2}} &= \frac{1}{\sqrt{2}}e^{\pm i \phi}\left[\sin \tau^+ \cos \tau^+\partial_+ -  \sin \tau^-\cos \tau^-\partial_-\pm i \sin(\tau^++ \tau^-)\csc(\tau^+-\tau^-) \partial_\phi\right].
\end{align}
which we will collectively denote $\zeta_A$, $A=1,\cdots 10$. Their Lie brackets are  
\begin{align}\label{Eqn: SO(3,2) commutators}
    \left[L_m,L_n\right]\hspace{20pt}&= -i(m-n)L_{m+n}, && \left[\bar{L}_{\bar{m}},\bar{L}_{\bar{n}}\right]\hspace{8pt} = -i(\bar{m}-\bar{n})\bar{L}_{\bar{m}+\bar{n}},\nonumber\\
    \left[L_m,H_{\bar{r},s}\right]\hspace{9pt} &= -i\left(\frac{m}{2}-s\right)H_{\bar{r},m+s}, && \left[\bar{L}_{\bar{m}},H_{\bar{r},s}\right] = -i\left(\frac{\bar{m}}{2}-\bar{r}\right)H_{\bar{m}+\bar{r},s},\nonumber\\
     \left[H_{\bar{r},s}, H_{\bar{r}',s'}\right] &= i(s-s')\bar{L}_{\bar{r}+\bar{r}'}+i(\bar{r}-\bar{r}')L_{s+s'},
\end{align}
for $m, \bar m,n,\bar n = 1,0,-1$ and $\bar r,\bar r',s,  s'= \pm \frac{1}{2}$. The $SO(3,2)$ action is generated  by the 10 charges
\be Q(\zeta_A)=\int_{S^2} d^2\Sigma^\mu j_{A\mu},~~~~  j_{A\mu}=T_{\mu\nu}\zeta^\nu_A.\label{eqn: SO(3,2) charge}\ee
Since $j_A$ is conserved, $S^2$ can be any Cauchy surface including $S^2_{x_i}$.  These charges act on the stress tensor as 
\begin{align}\label{eqn: T variation}
    \left[Q(\zeta_A),T_{\mu\nu}\right] 
     = -i(\mathcal{L}_{\zeta_A} T)_{\mu\nu}- \frac{i}{3}\nabla_\lambda\zeta_A^\lambda T_{\mu\nu},
\end{align}

From the definitions \eqref{eqn:lightray E EC}-\eqref{eqn: lightray N EC}  we find that zero modes of the Cordova-Shao light rays  give 3 of these 10 adjoint charges in \eqref{eqn: SO(3,2) charge}:
\begin{align}\label{mbssg}
Q( H_{-\half,-\half})&=-\sqrt{2}\mathcal{E}_0   \\
    Q(L_0)&=\half(\mathcal{N}_0+i\mathcal{K}_0)  \\
  Q(\bar L_0)&=-\half(\mathcal{N}_0-i\mathcal{K}_0).
    \end{align}

     Light transforms of any of the 10 conserved currents are
  \be\label{lj} L(\zeta_A;\phi)= \int_0^\pi d\tau^+ \sin \tau^+ j_{A+}(\tau^+,0,\phi).\ee
  Among these the ANEC plays a special role because it is annihilated by $Q(H_{-\half,-\half})$,  $Q(L_{-1})$ and $Q(\bar L_{-1})$ and so is  lowest  weight under $SO(3,2)$. Beyond \eqref{eqn:lightray E EC}-\eqref{eqn: lightray N EC}, \eqref{lj} gives seven operators
  \begin{align}
   L( L_{1};\phi) &=  \frac{i}{2}e^{ i\phi}\int_0^\pi d\tau^+\sin \tau^+\left[\cos^2 \tau^+ T_{++} -T_{+-} +2i\cot \tau^+T_{+\phi}\right],\\
 L(L_{-1};\phi) &=  ie^{ -i\phi}\mathcal{E}(\phi),\\
 L( \bar{L}_{1},;\phi) &=  \frac{i}{2}e^{-i\phi}\int_0^\pi d\tau^+\sin \tau^+\left[\cos^2 \tau^+ T_{++}-T_{+-} -2i\cot \tau^+T_{+\phi}\right],\\
    L(\bar L_{-1};\phi) &=  ie^{ i\phi}\mathcal{E}(\phi),\\
   L( H_{\frac{1}{2},\frac{1}{2}};\phi) &= -\frac{1}{\sqrt{2}}\int_0^\pi d\tau^+\sin \tau^+\left[\cos^2 \tau^+T_{++}+T_{+-}\right],\\
  L(H_{\mp\frac{1}{2},\pm\frac{1}{2}};\phi) &= \frac{1}{\sqrt{2}}e^{\pm i \phi}\bigl(-{\mathcal K}\pm i{\mathcal N}\bigr).
  \end{align}

  The factor of $\sin \tau^+=\sqrt{g_{\phi\phi}}|_{\tau^-=0}$ in \eqref{lj}  implies that it is a density as a function of $\phi$. 
 Fourier transforming in  $\phi$ then gives the smeared light ray operators  
     \be L(\zeta_A;k)= \int_0^{2\pi}d\phi e^{ik \phi} \int_0^\pi d\tau^+ \sin \tau^+ j_{A+}(\tau^+,0,\phi).\ee  
     Introducing  the notation 
     \bea \label{csm}  w^{3+k\over 2}_{\bar{r}-{k\over 2},s+{k\over 2}}&\equiv & {i^{-k}\over \sqrt{2}}L(H_{\bar{r},s};k),\cr
    w^{1+{k\over 2}}_{-{k\over 2},n+{k\over 2}}&\equiv &-i^{-k} L(L_n;k),\cr
     w^{2+{k\over 2}}_{\bar{n}-{k\over 2},{k\over 2}}&\equiv & i^{-k} L(\bar L_{\bar{n}};k),  \label{sax}
     \eea
     one finds
     \begin{align}\label{mbss1}
    \mathcal{E}_k &=-i^k w^{3+k\over 2}_{-{k+1\over 2},{k-1\over 2}},  \\
   \mathcal{N}_k+i\mathcal{K}_k& =-2i^k w^{1+{k\over 2}}_{-{k\over 2},{k\over 2}},\\
    \mathcal{N}_k-i\mathcal{K}_k& =-2i^k w^{2                   +{k\over 2}}_{-{k\over 2},{k\over 2}}.\end{align}
The  commutation relations \eqref{eqn: comm E E K K K E}, \eqref{eqn: comm N E N K N N}, and \eqref{Eqn: SO(3,2) commutators} comprise  a subset within the more general expression 
\begin{align}\label{eqn: deformed w}
    \left[w^p_{\bar{m},m},w^q_{\bar{n},n}\right] = (\bar{m}(q-1)-\bar{n}(p-1))w^{p+q-2}_{\bar{m}+\bar{n},m+n} +  (m(q-2)-n(p-2))w^{p+q-1}_{\bar{m}+\bar{n},m+n},
\end{align}
 which is the algebra \dw\  for cosmological constant $\Lambda=-1$~\cite{taylor2023,bittleston2024}. 

 The indices on the element $w^p_{\bar m , m}$ denote its transformation properties under the  $SO(2,2)$ subgroup of $SO(3,2)$ generated by ($\bar L_{\bar{n}},L_n$). $\bar m$
and $m$ are eigenvalues of the generators $\bar L_0, L_0$ of the Cartan subalgebra. For fixed $p$, $w^p_{\bar m , m}$ form a lowest weight $SO(2,2)$ representation (annihilated by $L_{-1},~\bar L_{-1}$) labeled by $(\bar h,  h)=(p,3-p)$.

ANEC modes $w^p_{1-p, p-2}$ are lowest weight  because 
\begin{align}
i^{1-2q}\left[w^p_{\bar{m},m},\mathcal{E}_{2q-3}\right]=\left[w^p_{\bar{m},m},w^q_{1-q,q-2}\right] &= (q-1)(\bar{m}+p-1)w^{p+q-2}_{\bar{m}+1-q,m+q-2}\nonumber\\&+(q-2)(m-p+2)w^{p+q-1}_{\bar{m}+1-q,m+q-2},
\end{align}
vanishes for the $SO(3,2)$ lowering operators which have $(p,\bar m , m)$ equal to $(1, 0,-1),~(2,-1,0), ~({3 \over 2},-\half,-\half)$.  These modes $w^p_{\bar m , m}$ assemble into a single $SO(3,2)$ representation with quadratic Casimir equal to 6.

\section{The Wedge }

    The (Fourier transformed) Cordova-Shao operators defined in \eqref{mbss}  do not define $w^p_{\bar m,m}$ for every value of ($p, \bar m, m$) with $p\pm m\in \mathbb{Z}$ and 
    $p\pm \bar m \in \mathbb{Z}$.   Moreover, they do not close under commutation with the $SO(3,2)$ generators in \eqref{eqn: SO(3,2) charge}. We  consider the closed algebra of all conformal transformations of the  operators in  \eqref{csm} and all commutators thereof. We wish to show that 
    
\begin{figure}
        \centering
        \includegraphics[width=140mm]{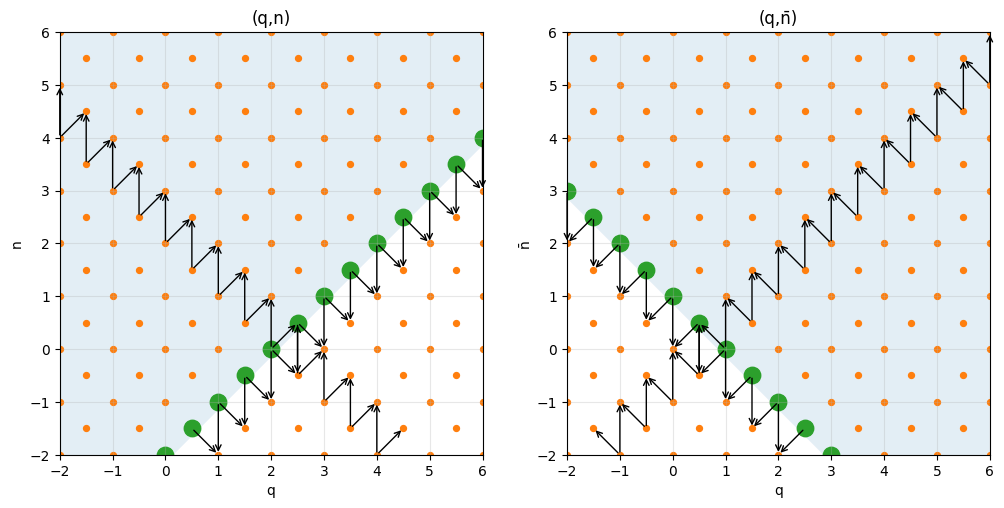}
        \caption{Lattice of states. The green dots  are  modes of ANEC operator.  These are lowest weight states at the edge of the wedge \eqref{swedge}, which is the shaded region.  The arrow is the ``forbidden" operations along which all the $SO(3,2)$ generators have vanishing  coefficients.}
        \label{fig: wedge}
    \end{figure}

    \noindent(i) this generates a unique operator with  every  value of ($p,  \bar m, m$) lying in the  wedge\footnote{This particular  wedge subalgebra has been previously encountered in \cite{bittleston2024}.} 
    \be\label{swedge}\bar{m}+p\geq 1,~~~~m-p\geq-2,\ee
    (see Fig.~\ref{fig: wedge})  and 
    
    \noindent (ii) the operators obey the commutation relations \eqref{eqn: deformed w}.

    Together with the facts that the Cordova-Shao operators are $SO(3,2)$ transformations of the ANEC operator, and the ANEC operator is $SO(3,2)$ lowest-weight,  (i) and (ii) imply that descendants of the ANEC generate the wedge \eqref{swedge} of \dw.

       To demonstrate (i) we first note that all the elements in \eqref{sax} are in the wedge. In \eqref{eqn: deformed w}, the right hand side has $m'-p'\geq m-p+{n}-q+1\geq -2$.\footnote{Note that when   $m-p =n-q=-2$,  the coefficient of the second term vanishes as $m(q-2)-n(p-2)=0$.} Similarly, the antiholomorphic index on the  right hand side is  in the wedge $\bar{m}+p \geq 1 $. Therefore, commutators of operators  in the wedge are themselves  in the wedge.  In the CFT$_3$ construction, the objects that saturate the wedge condition are the Fourier modes of the ANEC operator.

It remains to show that every operator in the wedge \eqref{swedge} can be reached by conformal transformations of the Cordova-Shao operators \eqref{sax}. In fact we need only the  lowest-weight ANEC operator \begin{align}
    \mathcal{E}_k = -i^k w^{\frac{3+k}{2}}_{-\frac{k+1}{2},\frac{k-1}{2}}.
\end{align} 
Setting $k=2q-3$ gives $w^q_{1-q,q-2}$, so any $q$ is clearly reachable. The $(\bar n, n)$ index can be raised and lowered using 
\begin{align}\label{eqn: L1, Lb1 action}
\left[Q(L_1),w^q_{\bar{n},n}\right] \hspace{11pt}&= \left(2-q-n\right)w^q_{\bar{n},n+1}, \hspace{20pt} 
    \left[Q(\bar{L}_{1}),w^q_{\bar{n},n}\right] = \left(q-1-\bar{n}\right)w^q_{\bar{n}+1,n},
\end{align}
\begin{align}\label{ction}
\left[Q(L_{-1}),w^q_{\bar{n},n}\right] \hspace{11pt}&= \left(q-2-n\right)w^q_{\bar{n},n-1}, \hspace{20pt} 
    \left[Q(\bar{L}_{-1}),w^q_{\bar{n},n}\right] = \left(1-q-\bar{n}\right)w^q_{\bar{n}-1,n},
\end{align}
to obtain  $some$ of the rest of the elements in this wedge. However obstacles are sometimes encountered at $n=2-q$ or $\bar{n}= q-1$ where the coefficent of the raising operation vanishes. Nevertheless it is always possible to go around such obstacles using other generators. The proof is given in Appendix~\ref{app: every object}. The structure of the wedge and forbidden transitions is illustrated in Fig. \ref{fig: wedge}.

Uniqueness follows from the fact that every operator can be lowered to the ANEC mode with the same value of $p$ and those are unique. (ii) then follows using the Jacobi identity. Constructive  proofs of these statements are in  Appendix C.

We note that operators outside the wedge may exist in CFT$_3$ but are not given by our construction. 

\section{AdS$_4$}

While this paper focuses on  the boundary picture, we  briefly comment here on the bulk AdS$_4$  picture. 

The \dw\ algebra was discovered   in \cite{taylor2023} by perturbing the flat space OPEs with  a cosmological constant interaction.  In \cite{bittleston2024} it was found  by deforming the twistorial description of self-dual gravity from  flat space to negative curvature. In both cases the \dw\ algebra is generated by positive helicity soft gravitons. Explicit expression for the self-dual graviton modes corresponding to $w^p_{\bar m, m}$ can be found in \cite{bittleston2024}. In neither case was the effect of AdS$_4$ boundary conditions considered. Indeed, the standard Dirichlet boundary conditions\footnote{Boundary conditions which allow purely self-dual field configurations do exist, albeit with some unusual but interesting properties \cite{Aharony:2024xxx,Jain_2025,Mishra_2019,skvortsov2026}. } reflect positive helicity to negative helicity and are not compatible with self-duality. 

A similar situation arose in \cite{Sheta:2025oep} in the context of nonabelian gauge theory. In flat space, there are soft $S$ and $\bar S$ algebra generators $S^{p,a}_{\bar m, m}$ and $\bar S^{p,a}_{\bar m, m}$ comprised of positive and negative helicity gluons. When conformally mapped to AdS$_4$, these modes violate Dirichlet boundary conditions. Nevertheless, it was shown that linear combinations of the form
\be T^{p,a}_{\bar m, m}=S^{p,a}_{\bar m, m}+\bar{S}^{p,a}_{-m,-\bar m }\ee 
are comprised of linearly polarized gluons and generate an algebra isomorphic to the $S$-algebra. These linear combinations  survive the $\mathbb{Z}_2$ quotient which produces AdS$_4$ from EC$^4$. 
We expect something similar here, perhaps involving  combinations of the form  ${w}^{p}_{\bar{m},m}\pm\bar{w}^{3-p}_{\bar m,{m}}$. 

A careful derivation of \dw\ from the bulk perspective incorporating AdS$_4$ boundary conditions would be of interest.

\section*{\Large\bf Acknowledgements}

We are grateful to Matthew Dodelson, Simon Hueveline, Shu-Heng Shao  and especially   Ahmed Sheta and Adam Tropper for many useful discussions.
This work was  supported by the Simons Collaboration for Celestial Holography, the Black Hole Initiative and NSF grant PHY–2207659. 
\appendix
\section{Conformal mappings EC$^3\leftrightarrow$M$^3$}\label{appendix: Coords Light Ray}
This appendix gives conformal transformations between the Einstein cylinder EC$^3$ with coordinates $(\tau^+,\tau^-,\phi)$ and Minkowski space with flat coordinates $ (y^+,y^-,y)$.
The metric, coordinates and stress tensor are related by  
\begin{align}
    (y^+,y^-,y) &= \Omega\left( -2 \cos \tau^+ \cos \tau^-,2\sin\tau^+ \sin\tau^-,\sin(\tau^+-\tau^-)  \sin \phi\right)\\
   ds^2_{\text{M}^3}&=\Omega^{2} ds^2_{\text{EC}^3},\\
    T_{\mu\nu}dx^\mu dx^\nu|_{\text{M}^3}&=\Omega^{-1}T_{\mu\nu}dx^\mu dx^\nu|_{\text{EC}^3},\\
    \Omega &= {1 \over \sin(\tau^++\tau^-)+\sin(\tau^+-\tau^-)\cos\phi}.
\end{align}
Using this, our EC$^3$  lightray expressions \eqref{eqn:lightray E EC}-\eqref{eqn: lightray N EC} become Cordova and Shao Lightrays (3.1)-(3.3) in reference ~\cite{Cordova_2018}: 
\begin{align}
    \mathcal{E}(y) = \mathcal{E}(\phi) \left(\frac{d\phi}{dy}\right)^2, \hspace{10pt}\mathcal{K}(y) = \mathcal{K}(\phi) \left(\frac{d\phi}{dy}\right), \hspace{10pt}\mathcal{N}(y) = \mathcal{N}(\phi) \left(\frac{d\phi}{dy}\right)^2 + \mathcal{K}(\phi)\left(\frac{d^2\phi}{dy^2}\right),\label{eqn: mapping from flat to EC}
\end{align}
where $y = \tan(\phi/2)$ in our choice of coordinates.
\section{Plane model algebra}

The wedge condition appearing in \eqref{swedge}
\be\label{rswedge}\bar{m}+p\geq 1,~~~~m-p\geq-2\ee is different  from the flat space wedge 
\be \label{fwedge} p = 1,\frac{3}{2},2,..., ~~~~~~\bar{m} = 1-p,...,p-1. \ee Redefining the generators 
\begin{align}
    w^{p}_{\bar{m},m} &= -\frac{1}{2}i^{m'+\bar{m}'+2p'}\hat{w}^{p'}_{\bar{m}',m'}\label{eqn: relation W}\\\left(p',\bar{m}',m'\right) &= \left(\frac{3}{2}+\frac{\bar{m}+m}{2},\;p-\frac{3}{2}+\frac{\bar{m}-m}{2},\;p-\frac{3}{2}-\frac{\bar{m}-m}{2}\right),\nonumber
\end{align}
one finds that  $\hat{w}^p_{\bar{m},m}$ lie  in the  wedge \eqref{fwedge}.  These obey the `plane-model' algebra 
\begin{align}\label{eqn: Plane model algebra}
\left[\hat{w}^p_{\bar{m},m},\hat{w}^q_{\bar{n},n}\right] &= \left[(p+\bar{m}-1)(q-n-2)-(q+\bar{n}-1)(p-m-2)\right]\hat{w}^{p+q-\frac{3}{2}}_{\bar{m}+\bar{n}-\frac{1}{2},m+n-\frac{1}{2}}\nonumber\\&+\left[(p-\bar{m}-1)(q+n-2)-(q-\bar{n}-1)(p+m-2)\right]\hat{w}^{p+q-\frac{3}{2}}_{\bar{m}+\bar{n}+\frac{1}{2},m+n+\frac{1}{2}}.
\end{align}
appearing in \cite{bittleston2024,Heuveline:2025nmb}.

\section{Constructive proof of the wedge algebra}\label{app: every object}
\begin{figure}
    \centering
    \includegraphics[width=150mm]{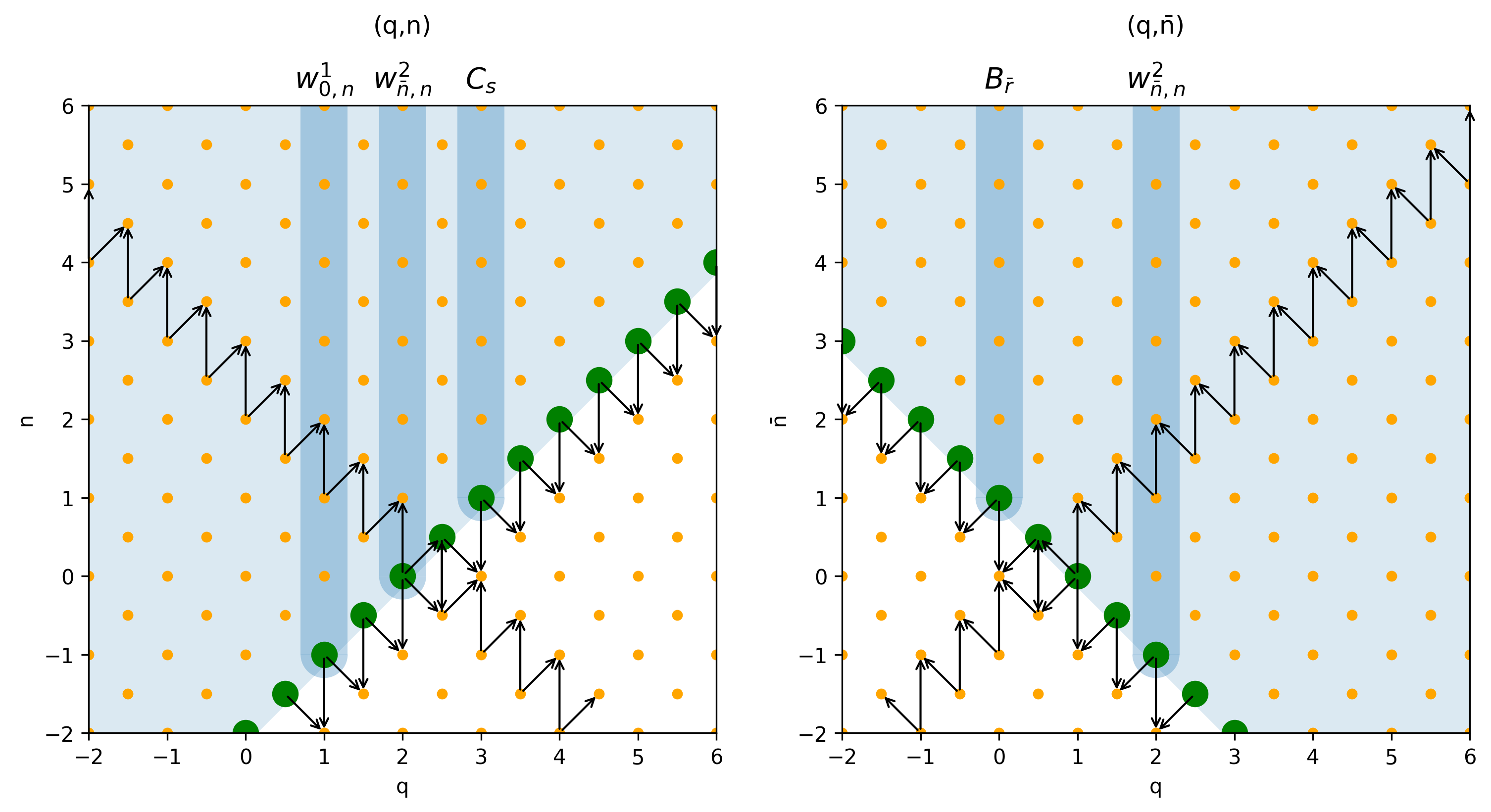}
    \caption{Objects used in the construction. Green dots denote the ANEC modes $A_q$. Other objects are indicated as vertical strips in the figure. The arrow is the ``forbidden" operations along which all the $SO(3,2)$ generators have vanishing  coefficients.}
    \label{fig:construction}
\end{figure}
This  appendix completes the argument   of section 4. 
We show by construction that every generator $w^p_{\bar m,m}$ in the wedge \eqref{swedge}
\begin{align}
\bar m+p\ge 1,
\qquad
m-p\ge -2,
\end{align}
is obtained from the ANEC modes
\begin{align}
A_q\equiv w^q_{1-q,q-2}.
\end{align}
We begin by defining 
\begin{align}
B_{\bar{r}}\equiv w^0_{{\bar{r}},-2},
\qquad
C_s\equiv w^3_{-2,s},
\qquad {\bar{r}},s\ge 1,
\end{align}
which are obtined by repeated action of $Q(\bar L_1)$ and $Q(L_1)$  from $A_0$ and $A_3$. There is no obstruction from vanishing coefficients,  see  figure \eqref{fig:construction}.
Next we define
\begin{align}
\,w^2_{-1,m}\equiv \frac{1}{2(m+1)}[B_{1},C_{m+2}],~~ m\geq 0.
\end{align}
Action by $Q(\bar L_1)$ defines all $w^2_{0,m}$, $m\geq 0$, noting that $w^2_{0,-1}= A_2$ is already defined as an ANEC mode. We then have
\begin{align}
\,w^1_{0,m}\equiv \frac{1}{2}[B_{2},C_{m+2}]-(m+1)w^2_{0,m},~~ m\geq 0,
\end{align}
where again $w^1_{0,-1}=  A_1$ is already an ANEC mode.

Next, we define the lower boundary of $m$ for $p\neq -1$
\begin{align}
\,w^p_{\bar{m},p-2}\equiv \frac{1}{(p+1)(\bar{m}+p)}[B_{\bar m+p+1},A_{p+2}],
\end{align}
and $\bar{m}+p\geq 1$ for objects in the wedge. The exceptional case $p=-1$ is already contained in the $Q(L_{1})$ descendants of $A_{-1}=w^{-1}_{2,-3}$, which defines all $w^{-1}_{\bar m,-3}$ with $\bar m\ge 2$.

Once the lower boundary is known, the interior follows. For $p\neq 2$,
\begin{align}
w^p_{\bar m,m}\equiv \frac{1}{(2-p)(m-p+3)}[w^p_{\bar m,p-2},w^1_{0,m-p+2}],
\end{align}
and $m-p+3> 0$ throughout the wedge. The remaining value of  $p=2$ is defined from
\begin{align}
w^2_{\bar m,m}\equiv \frac{1}{(\bar m+2)}[w^2_{\bar m+1,0},w^2_{-1,m}].
\end{align}
This gives a constructive definition of $w^p_{\bar m,m}$ starting from ANEC modes every lattice point in the wedge.

It remains to show that the commutators of the operators  $w^p_{\bar m,m}$ so defined are  the \dw\ algebra. Since $Q(L_0)$ and $Q(\bar L_0)$ are Cartan generators, $[w^p_{\bar m,m},w^q_{\bar n,n}]$ has definite weights $(\bar m+\bar n,m+n)$ and   $ [w^p_{\bar m,m},w^q_{\bar n,n}] \sim \sum_r c_rw^r_{\bar m+\bar n,m+n}$.  Repeated action of $Q(L_{-1})$ or $Q(\bar L_{-1})$ eventually annihilates the commutator on the left hand side. Using $r-(p+q)\in\mathbb Z$, this constrains  the possible values of $r$: 
\begin{align}
r=p+q-2,
\qquad
r=p+q-1.
\end{align}
Therefore
\begin{align}\label{eqn: Ansatz}
[w^p_{\bar m,m},w^q_{\bar n,n}]
=
A\,w^{p+q-2}_{\bar m+\bar n,m+n}
+
B\,w^{p+q-1}_{\bar m+\bar n,m+n}
\end{align}
for some coefficients $A$ and $B$.

We can lower \eqref{eqn: Ansatz} with $Q(L_{-1})$ or $Q(\bar L_{-1})$: Jacobi identity expresses left hand side in terms of commutators at lower level. Equating it with the lowered right hand side fixes $A$ and $B$ to be precisely as in the \dw\ algebra \eqref{eqn: deformed w}. Thus the wedge is generated by communtators of  descendants of the ANEC operators, and the resulting generators satisfy the \dw\  commutation relations.
\bibliography{references}

\providecommand{\href}[2]{#2}\begingroup\raggedright\begin{thebibliography}{10}

\bibitem{Guevara:2021abz}
A.~Guevara, E.~Himwich, M.~Pate, and A.~Strominger, ``{Holographic symmetry algebras for gauge theory and gravity},'' \href{http://dx.doi.org/10.1007/JHEP11(2021)152}{{\em JHEP} {\bfseries 11} (2021) 152}, \href{http://arxiv.org/abs/2103.03961}{{\ttfamily arXiv:2103.03961 [hep-th]}}.

\bibitem{Strominger:2021mtt}
A.~Strominger, ``{$w_{1+\infty}$ Algebra and the Celestial Sphere: Infinite Towers of Soft Graviton, Photon, and Gluon Symmetries},'' \href{http://dx.doi.org/10.1103/PhysRevLett.127.221601}{{\em Phys. Rev. Lett.} {\bfseries 127} no.~22, (2021) 221601}, \href{http://arxiv.org/abs/2105.14346}{{\ttfamily arXiv:2105.14346 [hep-th]}}.

\bibitem{Sheta:2025oep}
A.~Sheta, A.~Strominger, A.~Tropper, and H.~Wei, ``{Soft Algebras in AdS$_4$ from Light Ray Operators in CFT$_3$},'' 12, 2025.
\newblock \url{https://arxiv.org/abs/2601.00096}.

\bibitem{Ball:2021tmb}
A.~Ball, S.~A. Narayanan, J.~Salzer, and A.~Strominger, ``Perturbatively exact $w_{1+\infty}$ asymptotic symmetry of quantum self-dual gravity,'' \href{http://dx.doi.org/10.1007/JHEP01(2022)114}{{\em JHEP} {\bfseries 01} (2022) 114}, \href{http://arxiv.org/abs/2111.10392}{{\ttfamily arXiv:2111.10392 [hep-th]}}.

\bibitem{mason2023}
L.~Mason, ``Gravity from holomorphic discs and celestial $\mathcal{L}w_{1+\infty}$ symmetries,'' 2023.
\newblock \url{https://arxiv.org/abs/2212.10895}.

\bibitem{adamo2022}
T.~Adamo, L.~Mason, and A.~Sharma, ``Celestial $w_{1+\infty}$ symmetries from twistor space,'' \href{http://dx.doi.org/10.3842/sigma.2022.016}{{\em Symmetry, Integrability and Geometry: Methods and Applications} (March, 2022) }. \url{http://dx.doi.org/10.3842/SIGMA.2022.016}.

\bibitem{kmec2024}
A.~Kmec, L.~Mason, R.~Ruzziconi, and A.~Y. Srikant, ``Celestial $\mathcal{L}w_{1+\infty}$ charges from a twistor action,'' 2024.
\newblock \url{https://arxiv.org/abs/2407.04028}.

\bibitem{himwich2023}
E.~Himwich and M.~Pate, ``$w_{1+\infty}$ in 4d gravitational scattering,'' 2023.
\newblock \url{https://arxiv.org/abs/2312.08597}.

\bibitem{Miller:2025wpq}
N.~Miller, ``{Spacetime $\mathcal{L}w_{1+\infty}$ Symmetry and Self-Dual Gravity in Plebanski Gauge},'' \href{http://arxiv.org/abs/2504.07176}{{\ttfamily arXiv:2504.07176 [hep-th]}}.

\bibitem{Bu:2024wnf}
W.~Bu and S.~Seet, ``{2d theory for asymptotic dynamics of 4d (self-dual) Einstein gravity},'' \href{http://arxiv.org/abs/2412.00918}{{\ttfamily arXiv:2412.00918 [hep-th]}}.

\bibitem{Ruzziconi:2025fuy}
R.~Ruzziconi and C.~Zwikel, ``{Celestial $\mathcal{L}w_{1+\infty}$ symmetries and subleading phase space of null hypersurfaces},'' \href{http://dx.doi.org/https://doi.org/10.48550/arXiv.2511.07525}{{\em Phys. Rev. D} {\bfseries 113} no.~4, (2026) 044067}, \href{http://arxiv.org/abs/2511.07525}{{\ttfamily arXiv:2511.07525 [hep-th]}}.

\bibitem{Cresto:2024fhd}
N.~Cresto and L.~Freidel, ``{Asymptotic higher spin symmetries I: covariant wedge algebra in gravity},'' \href{http://dx.doi.org/10.1007/s11005-025-01921-4}{{\em Lett. Math. Phys.} {\bfseries 115} no.~2, (2025) 39}, \href{http://arxiv.org/abs/2409.12178}{{\ttfamily arXiv:2409.12178 [hep-th]}}.

\bibitem{Donnay:2024qwq}
L.~Donnay, L.~Freidel, and Y.~Herfray, ``{Carrollian $\mathcal{L}w_{1+\infty}$ representation from twistor space},'' \href{http://dx.doi.org/10.21468/SciPostPhys.17.4.118}{{\em SciPost Phys.} {\bfseries 17} no.~4, (2024) 118}, \href{http://arxiv.org/abs/2402.00688}{{\ttfamily arXiv:2402.00688 [hep-th]}}.

\bibitem{Schwarz2023SymmetriesCelestial}
J.~H. Schwarz, ``Symmetries in celestial holography.'' \url{https://simonscelestialholographycollaboration.org/kickoff-workshop-videos/}, Oct., 2023.
\newblock Talk given at the Kickoff Workshop of the Simons Collaboration on Celestial Holography, Cambridge, MA.

\bibitem{taylor2023}
T.~R. Taylor and B.~Zhu, ``$w_{1+\infty}$ algebra with a cosmological constant and the celestial sphere,'' \href{http://dx.doi.org/10.1103/PhysRevLett.132.221602}{{\em Phys. Rev. Lett.} {\bfseries 132} (May, 2024) 221602}. \url{https://link.aps.org/doi/10.1103/PhysRevLett.132.221602}.

\bibitem{bittleston2024}
R.~Bittleston, G.~Bogna, S.~Heuveline, A.~Kmec, L.~Mason, and D.~Skinner, ``On ads$_4$ deformations of celestial symmetries,'' 2024.
\newblock \url{https://arxiv.org/abs/2403.18011}.

\bibitem{Jorge2023}
J.~Mago, L.~Ren, A.~Yelleshpur~Srikant, and A.~Volovich, ``Deformed $w_{1+\infty}$ algebras in the celestial cft,'' \href{http://dx.doi.org/10.3842/sigma.2023.044}{{\em Symmetry, Integrability and Geometry: Methods and Applications} (July, 2023) }. \url{http://dx.doi.org/10.3842/SIGMA.2023.044}.

\bibitem{HartmanMathys2024LightRaySumRules}
T.~Hartman and G.~Mathys, ``Light-ray sum rules and the $c$-anomaly,'' {\em JHEP} {\bfseries 08} (2024) 008, \href{http://arxiv.org/abs/2405.10137}{{\ttfamily arXiv:2405.10137 [hep-th]}}.

\bibitem{HartmanMathys2023ANECRG}
T.~Hartman and G.~Mathys, ``Averaged null energy and the renormalization group,'' {\em JHEP} {\bfseries 12} (2023) 139, \href{http://arxiv.org/abs/2309.14409}{{\ttfamily arXiv:2309.14409 [hep-th]}}.

\bibitem{FaulknerLeighParrikarWang2016ModularANEC}
T.~Faulkner, R.~G. Leigh, O.~Parrikar, and H.~Wang, ``Modular hamiltonians for deformed half-spaces and the averaged null energy condition,'' {\em JHEP} {\bfseries 09} (2016) 038, \href{http://arxiv.org/abs/1605.08072}{{\ttfamily arXiv:1605.08072 [hep-th]}}.

\bibitem{BalakrishnanFaulknerKhandkerWang2019QNEC}
S.~Balakrishnan, T.~Faulkner, Z.~U. Khandker, and H.~Wang, ``A general proof of the quantum null energy condition,'' \href{http://dx.doi.org/10.1007/JHEP09(2019)020}{{\em JHEP} {\bfseries 09} (2019) 020}, \href{http://arxiv.org/abs/1706.09432}{{\ttfamily arXiv:1706.09432 [hep-th]}}.

\bibitem{HartmanKunduTajdini2017ANECFromCausality}
T.~Hartman, S.~Kundu, and A.~Tajdini, ``Averaged null energy condition from causality,'' {\em JHEP} {\bfseries 07} (2017) 066, \href{http://arxiv.org/abs/1610.05308}{{\ttfamily arXiv:1610.05308 [hep-th]}}.

\bibitem{CasiniTesteTorroba2017NullPlaneMarkov}
H.~Casini, E.~Teste, and G.~Torroba, ``Modular hamiltonians on the null plane and the markov property of the vacuum state,'' {\em J. Phys. A} {\bfseries 50} (2017) 364001, \href{http://arxiv.org/abs/1703.10656}{{\ttfamily arXiv:1703.10656 [hep-th]}}.

\bibitem{HofmanLiMeltzerPoland2016ColliderBounds}
D.~M. Hofman, D.~Li, D.~Meltzer, and D.~Poland, ``A proof of the conformal collider bounds,'' {\em JHEP} {\bfseries 06} (2016) 111, \href{http://arxiv.org/abs/1603.03771}{{\ttfamily arXiv:1603.03771 [hep-th]}}.

\bibitem{Kabat_2021}
D.~Kabat, G.~Lifschyt, P.~Nguyen, and D.~Sarkar, ``Light-ray moments as endpoint contributions to modular hamiltonians,'' \href{http://dx.doi.org/10.1007/jhep09(2021)074}{{\em Journal of High Energy Physics} {\bfseries 2021} no.~9, (Sept., 2021) }. \url{http://dx.doi.org/10.1007/JHEP09(2021)074}.

\bibitem{Rosso2020GlobalAspectsANEC}
F.~Rosso, ``Global aspects of conformal symmetry and the anec in ds and ads,'' \href{http://dx.doi.org/10.1007/JHEP03(2020)186}{{\em JHEP} {\bfseries 03} (2020) 186}, \href{http://arxiv.org/abs/1912.08897}{{\ttfamily arXiv:1912.08897 [hep-th]}}.

\bibitem{moult2025memorycorrelatorswardidentities}
I.~Moult, S.~A. Narayanan, and S.~Pasterski, ``Memory correlators and ward identities in the `in-in' formalism,'' 2025.
\newblock \url{https://arxiv.org/abs/2512.02825}.

\bibitem{Gonzalez:2025ene}
H.~A. Gonz{\'a}lez and J.~Salzer, ``{Energy Detectors and Asymptotic Symmetries},'' \href{http://arxiv.org/abs/2510.27348}{{\ttfamily arXiv:2510.27348 [hep-th]}}.

\bibitem{Cordova_2018}
C.~Córdova and S.-H. Shao, ``Light-ray operators and the bms algebra,'' \href{http://dx.doi.org/10.1103/physrevd.98.125015}{{\em Physical Review D} {\bfseries 98} no.~12, (Dec., 2018) }. \url{http://dx.doi.org/10.1103/PhysRevD.98.125015}.

\bibitem{Himwich:2025ekg}
E.~Himwich and M.~Pate, ``{Light-ray Operators and the $w_{1+\infty}$ Algebra},'' \href{http://arxiv.org/abs/2512.18973}{{\ttfamily arXiv:2512.18973 [hep-th]}}.

\bibitem{Hofman_2008}
D.~M. Hofman and J.~Maldacena, ``Conformal collider physics: energy and charge correlations,'' \href{http://dx.doi.org/10.1088/1126-6708/2008/05/012}{{\em Journal of High Energy Physics} {\bfseries 2008} no.~05, (May, 2008) 012}. \url{http://dx.doi.org/10.1088/1126-6708/2008/05/012}.

\bibitem{Kravchuk_2018}
P.~Kravchuk and D.~Simmons-Duffin, ``Light-ray operators in conformal field theory,'' \href{http://dx.doi.org/10.1007/jhep11(2018)102}{{\em Journal of High Energy Physics} {\bfseries 2018} no.~11, (Nov., 2018) }. \url{http://dx.doi.org/10.1007/JHEP11(2018)102}.

\bibitem{Karateev_2018}
D.~Karateev, P.~Kravchuk, and D.~Simmons-Duffin, ``Weight shifting operators and conformal blocks,'' \href{http://dx.doi.org/10.1007/jhep02(2018)081}{{\em Journal of High Energy Physics} {\bfseries 2018} no.~2, (Feb., 2018) }. \url{http://dx.doi.org/10.1007/JHEP02(2018)081}.

\bibitem{Kologlu:2019mfz}
M.~Kologlu, P.~Kravchuk, D.~Simmons-Duffin, and A.~Zhiboedov, ``{The light-ray OPE and conformal colliders},'' \href{http://dx.doi.org/10.1007/JHEP01(2021)128}{{\em JHEP} {\bfseries 01} (2021) 128}, \href{http://arxiv.org/abs/1905.01311}{{\ttfamily arXiv:1905.01311 [hep-th]}}.

\bibitem{Korchemsky:2021htm}
G.~P. Korchemsky and A.~Zhiboedov, ``{On the light-ray algebra in conformal field theories},'' \href{http://dx.doi.org/10.1007/JHEP02(2022)140}{{\em JHEP} {\bfseries 02} (2022) 140}, \href{http://arxiv.org/abs/2109.13269}{{\ttfamily arXiv:2109.13269 [hep-th]}}.

\bibitem{BelinHofmanMathysWalters2021StressTensorLightRay}
A.~Belin, D.~M. Hofman, G.~Mathys, and M.~T. Walters, ``On the stress tensor light-ray operator algebra,'' {\em JHEP} {\bfseries 05} (2021) 033, \href{http://arxiv.org/abs/2011.13862}{{\ttfamily arXiv:2011.13862 [hep-th]}}.

\bibitem{BeskenDeBoerMathys2021IntegratedCommutators}
M.~Be\c{s}ken, J.~de~Boer, and G.~Mathys, ``On local and integrated stress-tensor commutators,'' {\em JHEP} {\bfseries 07} (2021) 148, \href{http://arxiv.org/abs/2012.15724}{{\ttfamily arXiv:2012.15724 [hep-th]}}.

\bibitem{ChangKologluKravchukSimmonsDuffinZhiboedov2022TransverseSpin}
C.-H. Chang, M.~Kolo\u{g}lu, P.~Kravchuk, D.~Simmons-Duffin, and A.~Zhiboedov, ``Transverse spin in the light-ray ope,'' {\em JHEP} {\bfseries 05} (2022) 059, \href{http://arxiv.org/abs/2010.04726}{{\ttfamily arXiv:2010.04726 [hep-th]}}.

\bibitem{Huang2019NearLightcone}
K.-W. Huang, ``Stress-tensor commutators in conformal field theories near the lightcone,'' \href{http://dx.doi.org/10.1103/PhysRevD.100.061701}{{\em Phys. Rev. D} {\bfseries 100} (2019) 061701}, \href{http://arxiv.org/abs/1907.00599}{{\ttfamily arXiv:1907.00599 [hep-th]}}.

\bibitem{Huang_2020}
K.-W. Huang, ``Lightcone commutator and stress-tensor exchange in $d>2$ conformal field theories,'' \href{http://dx.doi.org/10.1103/physrevd.102.021701}{{\em Physical Review D} {\bfseries 102} no.~2, (July, 2020) }. \url{http://dx.doi.org/10.1103/PhysRevD.102.021701}.

\bibitem{Huang2021NearLine}
K.-W. Huang, ``$d>2$ stress-tensor operator product expansion near a line,'' {\em Phys. Rev. D} {\bfseries 103} (2021) 121702, \href{http://arxiv.org/abs/2103.09930}{{\ttfamily arXiv:2103.09930 [hep-th]}}.

\bibitem{HuPasterski2023CelestialConformalColliders}
Y.~Hu and S.~Pasterski, ``Celestial conformal colliders,'' {\em JHEP} {\bfseries 02} (2023) 243, \href{http://arxiv.org/abs/2211.14287}{{\ttfamily arXiv:2211.14287 [hep-th]}}.

\bibitem{HuPasterski2023DetectorOperators}
Y.~Hu and S.~Pasterski, ``Detector operators for celestial symmetries,'' \href{http://dx.doi.org/10.1007/JHEP12(2023)035}{{\em JHEP} {\bfseries 12} (2023) 035}, \href{http://arxiv.org/abs/2307.16801}{{\ttfamily arXiv:2307.16801 [hep-th]}}.

\bibitem{GonzoPokraka2021Detectors}
R.~Gonzo and A.~Pokraka, ``Light-ray operators, detectors and gravitational event shapes,'' {\em JHEP} {\bfseries 05} (2021) 015, \href{http://arxiv.org/abs/2012.01406}{{\ttfamily arXiv:2012.01406 [hep-th]}}.

\bibitem{Sharma2022AmbidextrousLightTransforms}
A.~Sharma, ``Ambidextrous light transforms for celestial amplitudes,'' {\em JHEP} {\bfseries 01} (2022) 031, \href{http://arxiv.org/abs/2107.06250}{{\ttfamily arXiv:2107.06250 [hep-th]}}.

\bibitem{BanerjeeBasuBhatkar2023LightTransformedGluon}
S.~Banerjee, R.~Basu, and S.~A. Bhatkar, ``Light transformed gluon correlators in ccft,'' {\em JHEP} {\bfseries 01} (2023) 075, \href{http://arxiv.org/abs/2203.06657}{{\ttfamily arXiv:2203.06657 [hep-th]}}.

\bibitem{HuEtAl2022FourPointLightRay}
Y.~Hu, L.~Lippstreu, M.~Spradlin, A.~Yelleshpur~Srikant, and A.~Volovich, ``Four-point correlators of light-ray operators in ccft,'' \href{http://dx.doi.org/10.1007/JHEP07(2022)104}{{\em JHEP} {\bfseries 07} (2022) 104}, \href{http://arxiv.org/abs/2203.04255}{{\ttfamily arXiv:2203.04255 [hep-th]}}.

\bibitem{Korchemsky_2022}
G.~P. Korchemsky, E.~Sokatchev, and A.~Zhiboedov, ``Generalizing event shapes: in search of lost collider time,'' \href{http://dx.doi.org/10.1007/jhep08(2022)188}{{\em Journal of High Energy Physics} {\bfseries 2022} no.~8, (Aug., 2022) }. \url{http://dx.doi.org/10.1007/JHEP08(2022)188}.

\bibitem{DeHuYelleshpurVolovich2022FourLightRay}
S.~De, Y.~Hu, A.~Yelleshpur~Srikant, and A.~Volovich, ``Correlators of four light-ray operators in ccft,'' {\em JHEP} {\bfseries 10} (2022) 170, \href{http://arxiv.org/abs/2206.08875}{{\ttfamily arXiv:2206.08875 [hep-th]}}.

\bibitem{BalakrishnanChandrasekaranFaulknerLevineShahbaziMoghaddam2022ReplicaDefects}
S.~Balakrishnan, V.~Chandrasekaran, T.~Faulkner, A.~Levine, and A.~Shahbazi-Moghaddam, ``Entropy variations and light ray operators from replica defects,'' \href{http://dx.doi.org/10.1007/JHEP09(2022)217}{{\em JHEP} {\bfseries 09} (2022) 217}, \href{http://arxiv.org/abs/1906.08274}{{\ttfamily arXiv:1906.08274 [hep-th]}}.

\bibitem{FreivogelStoffels2025PositivityLightRay}
B.~W. Freivogel and H.~Stoffels, ``On the positivity of light-ray operators,'' \href{http://arxiv.org/abs/2501.08386}{{\ttfamily arXiv:2501.08386 [hep-th]}}.

\bibitem{Aharony:2024xxx}
O.~Aharony, R.~R. Kalloor, and T.~Kukolj, ``A chiral limit for chern-simons-matter theories,'' \href{http://dx.doi.org/10.1007/JHEP10(2024)051}{{\em JHEP} {\bfseries 10} (2024) 051}, \href{http://arxiv.org/abs/2405.01647}{{\ttfamily arXiv:2405.01647 [hep-th]}}.

\bibitem{Jain_2025}
S.~Jain, D.~K.~S, and E.~Skvortsov, ``Hidden sectors of chern-simons matter theories and exact holography,'' \href{http://dx.doi.org/10.1103/physrevd.111.106017}{{\em Physical Review D} {\bfseries 111} no.~10, (May, 2025) }. \url{http://dx.doi.org/10.1103/PhysRevD.111.106017}.

\bibitem{Mishra_2019}
R.~K. Mishra, A.~Mohd, and R.~Sundrum, ``Ads asymptotic symmetries from cft mirrors,'' \href{http://dx.doi.org/10.1007/jhep03(2019)017}{{\em Journal of High Energy Physics} {\bfseries 2019} no.~3, (Mar., 2019) }. \url{http://dx.doi.org/10.1007/JHEP03(2019)017}.

\bibitem{skvortsov2026}
E.~Skvortsov and R.~V. Dongen, ``Dirichlet, neumann, mixed and self-dual holography: (self-dual) yang-mills theory,'' 2026.
\newblock \url{https://arxiv.org/abs/2602.21658}.

\bibitem{Heuveline:2025nmb}
S.~Heuveline, \href{http://dx.doi.org/10.17863/CAM.120155}{{\em {Celestial Chiral Algebras and Self-Dual Gravity}}}.
\newblock PhD thesis, Department of Applied Mathematics and Theoretical Physics, Cambridge U., Cambridge U. (main), 2025.
\newblock \href{http://arxiv.org/abs/2507.00772}{{\ttfamily arXiv:2507.00772 [hep-th]}}.

\end{thebibliography}\endgroup
\bibliographystyle{utphys}
\end{document}